\documentclass[a4paper,10pt]{article}
\usepackage[utf8]{inputenc}
\usepackage{fontenc}
\usepackage{graphicx}
\usepackage{bbm}
\usepackage [centertags] {amsmath}
\usepackage {amsfonts}
\usepackage{amsmath}
\usepackage{amssymb}
\usepackage {amsthm}
\usepackage {amscd}
\usepackage{cite}

\newtheorem{theorem}{Theorem}

\newtheorem{corollary}[theorem]{Corollary}

\newtheorem{proposition}[theorem]{Proposition}

\newcommand{\fml}[1]{(\ref{fml.#1})}

\newcommand{\seq}[3]{\left\{#1\right\}_{#2}^{#3}}
\newcommand{\SEQ}[3]{$\left\{#1\right\}_{#2}^{#3} $}
\newcommand{\fnc}[2]{#1\left(#2\right)}

\newcommand{\id}{\mathbbm{1}}

\newcommand{\setC}{\mathbbm{C}}
\newcommand{\setR}{\mathbbm{R}}
\newcommand{\setN}{\mathbbm{N}}
\def\Tr{\mathop{\textnormal{Tr}}}

\newcommand{\matrfour}[1]{\left[\begin{array}{cccc}#1_{00,00} & #1_{00,01} & #1_{00,10} & #1_{00,11}\\#1_{01,00} & #1_{01,01} & #1_{01,10} & #1_{01,11}\\#1_{10,00} & #1_{10,01} & #1_{10,10} & #1_{10,11}\\#1_{11,00} & #1_{11,01} & #1_{11,10} & #1_{11,11}\end{array}\right]}

\newcommand{\Repart}[1]{\fnc{\textnormal{Re}}{#1}}
\newcommand{\Impart}[1]{\fnc{\textnormal{Im}}{#1}}
\newcommand{\haskip}{\hskip 3 mm}
\newcommand{\Abs}[1]{\left|#1\right|}
\def\linspone{\mathcal H_1}
\def\linsptwo{\mathcal H_2}

\title{Positive maps, positive polynomials and entanglement witnesses}
\author{{\L}ukasz Skowronek$^{1}$ and Karol {\.Z}yczkowski$^{1,2}$\smallskip\\
$^1${\small Institute of Physics, Jagiellonian University, Krak{\'o}w, Poland}\\
$^2${\small Center for Theoretical Physics, Polish Academy of Sciences, Warsaw,
Poland}\\
{\small e-mail: lukasz.skowronek@uj.edu.pl karol@tatry.if.uj.edu.pl}\\
}

\begin{document}

\smallskipamount=2 mm

\maketitle

\begin{abstract}
We link the study of positive quantum maps, block positive operators,
and entanglement witnesses with problems related 
to multivariate polynomials. For instance, we show how indecomposable 
block positive operators relate to biquadratic forms that are not sums of squares.
Although the general problem of describing the set of positive maps 
remains open, in some particular cases we solve the corresponding 
polynomial inequalities and obtain explicit conditions for positivity.
\end{abstract}

\section{Introduction}

The set of positive maps acting on a finite dimensional Hilbert space
is a long-standing subject of mathematical interest.
In spite of many efforts (see \cite{StormerWor,Jamiolkowski,WorStormer,MM01,SSZ09} and references therein),
the structure of this set 
in spaces of arbitrary dimension is still not well understood. 
Of particular interest are positive maps,
which are not completely positive \cite{Choi75,Tomiyama,HaAtomic}.
The theorem of Jamio{\l}kowski implies \cite{Jamiolkowski} 
that any such map can be represented by an operator,
acting on a bi-partite Hilbert space, which is not positive, but is block-positive.

Non completely positive maps recently attracted a considerable attention of the physics community \cite{Sudarshan,Pachukas,Buzek}. Positive maps have mainly been studied
in view of their possible application to characterize quantum entanglement \cite{Horodeccy}
and in connection to entanglement witnesses 
\cite{TerhalIndec,LewensteinKraus, ChruscinskiKossakowski1,ChruscinskiKossakowski2}.
An entanglement witness is a Hermitian operator $W$ such that 
$\Tr\left(W\sigma\right)\geqslant 0$ for any separable state $\sigma$,
while negativity of $\Tr\left(W\rho\right)$ implies that the state $\rho$ is entangled.
Note that a Hermitian operator  $W$ may be considered as an
observable, so the expectation value $\Tr\left(W\rho\right)$ can be measured in 
an experiment \cite{GHB+03}.
From a mathematical perspective any entanglement
witness is a block positive operator which is not positive. 

In the present paper we aim to clarify the relation between positive maps and positive polynomials. 
Definitions and basic information can be found in Section \ref{sec.defs.blockpos}. 
In Section \ref{sectestingBP}, we explore the link between positive maps and positive polynomials 
and we address problems related to early contributions on the subject.
In particular, we analyze implications of the work of Jamio{\l}kowski \cite{Jamiolkowski1,Jamiolkowski2}
and show why the results of these papers do not allow one to formulate a conclusive test for positivity of a given map.

On the other hand, in some particular cases such results can be obtained.
In Sections \ref{secabc} and \ref{sec4x4test}, we investigate two families of maps
and working with the corresponding polynomials we find 
explicit conditions for positivity. Furthermore, we demonstrate how positive maps relate to the existence 
of positive polynomials which are not sums of squares and we formulate 
an open problem concerning entanglement witnesses in $2\times m$ dimensional spaces.

%%%%%%%%%%%%%%%%%%%%  section 2  %%%%%%%%%%%%%%%%%%%

\section{Block positivity - motivation and definitions}
\label{sec.defs.blockpos}

Let $\linspone$, $\linsptwo$ be finite dimensional spaces over $\setC$, both equipped with Hermitian inner products 
($\dim\linspone=N_1$, $\dim\linsptwo=N_2$). Let $\mathcal{L}\left(\linspone\right)$ denote the algebra of linear 
operators on $\linspone$. We denote with $\mathcal{L}\left(\linspone\right)^+$ the set of positive elements of 
$\mathcal{L}\left(\linspone\right)$. A linear map $\Phi:\mathcal{L}\left(\linspone\right)\rightarrow\mathcal{L}\left(\linsptwo\right)$ 
is called \textit{positive} if and only if it maps elements of $\mathcal{L}\left(\linspone\right)^+$  to elements of 
$\mathcal{L}\left(\linsptwo\right)^+$.
It is well known \cite{Jamiolkowski} that the set of positive maps is isomorphic to 
the set of \textit{block positive operators} (block positive over $\setC$). Therefore,   
instead of asking whether a given map is positive, in this work  we will be concerned with the
equivalent question whether the corresponding operator is block positive,
so it can serve as an entanglement witness.

A Hermitian operator $A$ on ${\cal H}=\linspone\otimes\linsptwo$ 
is called \textit{block positive over $\setC$} if it satisfies the following condition,
\begin{equation}
\label{blockposCdef}
 \left<u\otimes v\right|\left.A\left(u\otimes v\right)\right>
\geqslant 0\hskip 3 mm\forall_{u\in \linspone, v\in \linsptwo}.
\end{equation}
Note that condition \eqref{blockposCdef} is not invariant with respect to global unitary transformations on 
$\cal H$, so this definition depends on the particular form of the decomposition 
of $\cal H$.

It will also be useful to introduce the concept of block positivity for real linear spaces. 
Let $X$ and $Y$ be finite dimensional vector spaces over $\setR$ ($\dim X=M_1$, $\dim Y=M_2$). 
Let $A$ be a linear operator on $X\otimes Y$. In analogy to \eqref{blockposCdef}, we say that $A$ is 
\textit{block positive over $\setR$} if it satisfies
\begin{equation}
  \label{blockposRdef}
  \left(x\otimes y\right)\cdot A\left(x\otimes y\right)\geqslant 0\hskip 3 mm\forall_{x\in X, y\in Y}.
\end{equation}
Condition \eqref{blockposRdef} does not imply symmetry of $A$, but we may always assume that $A$ is symmetric because 
  the antisymmetric part of $A$  in \eqref{blockposRdef} vanishes. 
 Thence $\left(X\otimes Y\right)^2\ni\left(w_1,w_2\right)\mapsto w_1\cdot A\left(w_2\right)\in\setR$ is a symmetric bilinear 
form on $X\otimes Y$. 

In index notation, condition \eqref{blockposRdef} reads
\begin{equation}
\label{blockposdefRind}
A_{ab,cd}x^a y^b x^c y^d\geqslant 0\haskip\forall_{\seq{x^a}{a=1}{M_1},\seq{y^b}{b=1}{M_2}\subset\setR},
\end{equation}
where $x^a$ and $y^b$ are the coordinates of $x$ and $y$ with respect to the orthonormal bases 
\SEQ{e_i}{i=1}{M_1}, \SEQ{f_j}{j=1}{M_2} of $X$, $Y$ (resp.) which we are using.

Obviously, \eqref{blockposdefRind} is a positivity condition for a real multivariate polynomial of degree $4$.
 If the polynomial $A_{ab,cd}x^a y^b x^c y^d$ is a \textit{sum of squares}
   (SOS) of some other polynomials $P_i$, then we must have
\begin{equation}
  \label{SOS}
  A_{ab,cd}x^a y^b x^c y^d \ =\ \sum_iP_i^2 \ = \  \sum_i\left(B^i_{ab}x^ay^b\right)^2,
\end{equation}
where the real coefficients $B^i_{ab}$ ($a=1,\ldots,M_1$, $b=1,\ldots,M_2$) are arbitrary and the range of the index $i$ is finite.

Indeed, the polynomials $P_i$ must be homogeneous and of degree $2$. They cannot have terms of the form $x^ax^b$, 
neither of the form $y^ay^b$, since there are no terms $\left(x^ax^b\right)^2$ nor $\left(y^ay^b\right)^2$ in the 
sum $A_{ab,cd}x^a y^b x^c y^d$. Thus we conclude that if $A_{ab,cd}x^a y^b x^c y^d=\sum_iP_i^2$ for some polynomials $P_i$, 
then $P_i=B^i_{ab}x^ay^b$. But \eqref{SOS} looks just like a quadratic form on $X\otimes Y$, written in the product 
basis \SEQ{e_i\otimes f_j}{i=1,j=1}{M_1,M_2}. It is tempting to say that \eqref{SOS} implies positive semidefinitness 
of $A$, but \emph{this is not true}. Nevertheless, a similar result can be proved if we assume that $A$ is 
\textit{symmetric with respect to partial transpose}, $A^{\tau}:=\left(\id\otimes T\right)A=A$, where
$T$ denotes the transposition. 
Putting this in a different way, $A$ should satisfy
\begin{equation}
  \label{partTA}
  \left(x_1\otimes y_1\right)\cdot A\left(x_2\otimes y_2\right)= 
  \left(x_1\otimes y_2\right)\cdot A\left(x_2\otimes y_1\right)\haskip\forall_{x_1,x_2\in X,\,y_1,y_2\in Y}.
\end{equation}

For any operator $A$ being a SOS and expressed by eq.\,\eqref{SOS},
we may define the following operator $\tilde A$,
\begin{equation}
 \label{Atildematr}
 \tilde A_{ab,cd}=\frac{1}{2}\left(\sum_iB^i_{ab}B^i_{cd}+B^i_{ad}B^i_{cb}\right).
\end{equation}
It is easy to see that $\left(x\otimes y\right)\cdot\tilde A\left(x\otimes y\right)=\left(x\otimes y\right)\cdot
 A\left(x\otimes y\right)$ for all $x\in X$, $y\in Y$. In Appendix \ref{apphiA} we show that this property together 
with \eqref{partTA} and \eqref{Atildematr} imply $\tilde A=A$. But $\tilde A$ is of the special form \eqref{Atildematr}, 
which we did not assume about $A$. More precisely, $\tilde A$ is proportional to a sum of a semipositive 
 definite operator $B$ with matrix elements $\sum_iB^i_{ab}B^i_{cd}$ and its partial transposition $B^{\tau}$ with matrix elements $\sum_iB^i_{ad}B^i_{cb}$. We conclude that $A_{ab,cd}x^a y^b x^c y^d=\sum_iP_i^2$ implies
\begin{equation}
 \label{decomposbl}
 A=\frac{1}{2}\left(B+B^{\tau}\right),\haskip B\geqslant 0.
\end{equation}
for the operators $A$ with the property \eqref{partTA}. A Hermitian operator $A$ is called
\textit{decomposable} \cite{StormerWor,WorStormer}
iff $A=C+D^{\tau}$, where $C,D\geqslant 0$. When \eqref{partTA} holds, one can easily prove that
 \eqref{decomposbl} is equivalent to decomposability of $A$ . Thus we arrive at the following conclusion,
\begin{proposition}\label{indecompprop}
Let $X,Y$ be finite dimensional linear spaces over $\setR$. Let $\mathcal W$ be the set of 
\textbf{block positive, indecomposable} operators on $X\otimes Y$ which are symmetric with respect to
 transposition and \textbf{partial transposition}. Denote with $\mathcal P$ the set of 
\textbf{positive real polynomials} of the form $ A_{ab,cd}x^ay^bx^cy^d$ which are \textbf{not SOS}. 
There is a linear isomorphism between $\mathcal W$ and $\mathcal P$.

\begin{proof}
The isomorphism in question is $\Pi:{\mathcal W}\ni A\mapsto A_{ab,cd}x^a y^b x^c y^d\in{\mathcal P}$.
We still need to show that $\Pi$ is one-to-one. To this end, we assume the equality 
$\sum_{a,b,c,d} A_{ab,cd}x^ay^bx^cy^d=\sum_{a,b,c,d} B_{ab,cd}x^ay^bx^cy^d$ for some operators 
$ A, B\in\mathcal{W}$. Choose some $a,c\in\left\{1,2,\ldots,M_1\right\}$, $b,d\in\left\{1,2,\ldots,M_2\right\}$. 
Considering the coefficients at $x^ay^bx^cy^d$ in the two polynomials, we get $A_{ab,cd}+A_{ad,cb}= B_{ab,cd}+ B_{ad,cb}$.
 Thanks to the partial transpose symmetry of $A$ and $B$, we get $A=B$. This tells us that $\Pi$ is injective. 
On the other hand, every polynomial of the form $A_{ab,cd}x^ay^bx^cy^d$ is an image by $\Pi$ of the partial transpose 
symmetric operator $\frac{1}{2}\left(A+A^{\tau}\right)$. The operator $\frac{1}{2}\left(A+A^{\tau}\right)$ must be 
an element of $\mathcal W$ for $A_{ab,cd}x^ay^bx^cy^d$ to be an element of $\mathcal P$ (cf. the discussion above). 
Thus we conclude that $\Pi$ is surjective.
\end{proof}
\end{proposition}

It was demonstrated by Choi \cite{Choi75}  and St{\o}rmer \cite{Stormer}
 that there exist positive maps which are not 
decomposable. 
The example by Choi can be easily used to show that there exist, by Proposition \ref{indecompprop}, positive 
polynomials of the form $ A_{ab,cd}x^ay^bx^cy^d$ which are not SOS \cite{Choi75}.
Proposition \ref{indecompprop} gives a general motivation to investigate block positive operators 
over $\setR$ on account of their connection to sums of squares. It may also be expedient to study the real case 
in order to develop intuitions about block positivity over $\setC$. It should, however, be kept in mind that
 \eqref{blockposCdef} and \eqref{blockposRdef} \emph{are not the same}. Despite an apparent similarity, 
the block positivity over $\mathbbm C$ should not be perceived as a simple generalization 
of the block positivity over $\setR$.   
In general both definitions of block positivity do not coincide, 
what can be demonstrated by the following example of a real symmetric matrix,

\begin{equation}\label{exmplbpCnebpR}
 A=\left[\begin{array}{cccc}
               	1  &0 &  0         &-\frac{1}{2}\\
		0  &1 &\frac{3}{2} &0\\
		0& \frac{3}{2}  &1 &0\\
	-\frac{1}{2} &0      &0     &1\\
              \end{array}\right].
\end{equation}
This matrix represents an operator on $\setC^2\otimes\setC^2$ written in the standard product basis,
 $\{ |00\rangle, |01\rangle,|10\rangle,|11\rangle\}$.
It is easy to show that $A$ satisfies inequality \eqref{blockposRdef}, 
but it does not satisfy condition \eqref{blockposCdef}. Hence the matrix $A$ in \eqref{exmplbpCnebpR} is
block positive over $\setR$, 
but is not block positive over $\setC$. Moreover, when considering operators with unit trace, one can easily show that the set of such block positive operators over $\setC$ is compact whereas block positivity over $\setR$ does not imply compactness. In spite of this basic difference between the two notions of block positivity, there exist families of matrices for which conditions 
\eqref{blockposCdef} and \eqref{blockposRdef} turn out to be equivalent -- see Section \ref{secabc}.

%%%%%%%%%%%%%%%%%%%%  section 3  %%%%%%%%%%%%%%%%%%%

\section{Block positivity and quantifier elimination}
\label{sectestingBP}

Although the block positivity condition \eqref{blockposCdef} is simple to understand, it does not seem easy to check. 
The early papers by Jamio{\l}kowski \cite{Jamiolkowski1,Jamiolkowski2} suggest that the problem can be solved effectively.
Even though this conclusion is in some sense true, we show a weak point of the argument presented in these papers.

 For convenience of the reader, let us recall the details of the reasoning presented in \cite{Jamiolkowski1}. 
First, we write condition \eqref{blockposCdef} in index notation,
\begin{equation}
 \label{BPcondindex}
   A_{\alpha\beta,\gamma\delta}\bar u^{\alpha}\bar v^{\beta}u^{\gamma}v^{\delta}\geqslant 0 
   \haskip\forall_{\seq{u^{\alpha}}{\alpha=1}{N_1},\seq{v^{\beta}}{\beta=1}{N_2}\subset\setC}.
\end{equation}
Next, we introduce \textit{blocks}, 
\begin{equation}
 \label{blocks}
(A^{\left(1\right)}_v)_{\alpha\gamma}\ :=\ A_{\alpha\beta,\gamma\delta}\bar v^{\beta}v^{\delta},
\quad \quad
(A^{\left(2\right)}_u)_{\beta\delta}\ := \ A_{\alpha\beta,\gamma\delta}\bar u^{\alpha}u^{\gamma}. 
\end{equation}
We can interpret them simply as matrices or as operators on $\linspone$ and $\linsptwo$, respectively. 
Block positivity condition \eqref{BPcondindex} can be rewritten as
\begin{equation}
 \label{blockspositive}
 A^{\left(1\right)}_v\geqslant 0\haskip\forall_{v\in\linsptwo}\textrm{\haskip or 
as \haskip} A^{\left(2\right)}_u\geqslant 0\haskip\forall_{u\in\linspone},
\end{equation}
where ``$\geqslant$'' refers to semipositive definiteness. We shall concentrate on the right hand side of \eqref{blockspositive}.
 Semipositivity of $A^{\left(2\right)}_u$ is equivalent to the following set of inequalities,
\begin{eqnarray}
\label{minorspos}
 W_l\left(u\right):=\sum_{1\leqslant i_1<i_2<\ldots<i_l\leqslant N_2}\Delta_{i_1i_2\ldots i_l}\left(A^{\left(2\right)}_u\right)
\geqslant 0\haskip\forall_{u\in \linspone}\forall_{l=1\ldots N_2},
\end{eqnarray}
where $\Delta_{i_1i_2\ldots i_l}\left(A^{\left(2\right)}_u\right)$ is the minor of $A^{\left(2\right)}_u$ 
involving the columns and the rows with the numbers $i_1,\ldots,i_l$. It follows from the discussion in \cite{Jamiolkowski1} 
that the functions $W_l$ are homogeneous real polynomials of an even degree in the variables 
$\seq{\Repart{u^{\alpha}}}{\alpha=1}{N_1}$, $\seq{\Impart{u^{\gamma}}}{\gamma=1}{N_1}$. 
Thus \eqref{minorspos} is a set of positivity conditions for real homogeneous polynomials of an even degree. 
If we could solve these conditions explicitly, we would answer the question whether a given matrix is block positive.

That was the idea  presented in \cite{Jamiolkowski1} by  Jamio{\l}kowski, who suggested considering
 $\sum_{i_1,i_2,\ldots,i_n}C_{i_1i_2\ldots i_n}X_1^{i_1}X_2^{i_2}\ldots X^{i_n}_n$ ($C_{i_1i_2\ldots i_n}\in\setR$)
 as a polynomial in the variable $X_n$ with coefficients in $\setR\left[X_1,\ldots,X_{n-1}\right]$.
 He obtained positivity conditions for such a polynomial in a disjunctive normal form,
\begin{equation}
   \label{positivitycondDNF1}
   \forall_{\left\{x_1,\ldots,x_{n-1}\right\}\subset\setR}\bigvee_i\bigwedge_jD^i_j\left(x_1,\ldots,x_{n-1}\right)\geqslant 0,
\end{equation}
where $D^i_j\in\setR\left[X_1,\ldots,X_{n-1}\right]\forall_{i,j}$.
 Because the same procedure could be applied to any of the $D^i_j$'s 
(considered as elements of $\setR\left[X_1,\ldots,X_{n-2}\right]\left[X_{n-1}\right]$), 
it was claimed that the number of variables in \eqref{positivitycondDNF1} can be iteratively reduced so as to 
yield quantifier free formulas. The problem with this argument is that eq.\,\eqref{positivitycondDNF1} 
does not turn out to be equivalent to
\begin{equation}\label{positivitycondDNF2}
 \bigvee_i\bigwedge_j\forall_{\left\{x_1,\ldots,x_{n-1}\right\}\subset\setR}
\  D^i_j\left(x_1,\ldots,x_{n-1}\right)\geqslant 0,
\end{equation}
so one cannot use the procedure iteratively. 

To the best of our knowledge, no simple method  is known to check positivity 
of a general multivariate polynomial. It is in principle possible to eliminate quantifiers \cite{MichauxOzturk} from formulas like 
$\forall_{\left\{x_1,\ldots,x_{n-1}\right\}\subset\setR}\sum_{i_1,i_2,\ldots,i_n}C_{i_1i_2\ldots i_n}x_1^{i_1}\ldots x^{i_n}_n\geqslant 0$,
 but the outcome involves 
\emph{zeros of univariate polynomials of an arbitrary high degree}, which cannot in general be expressed in terms of the coefficients 
of the polynomials. The known quantifier elimination procedures are laborious and should not be expected to provide a constructive solution to the problem. 
Thus we have to conclude this section by repeating the accepted statement
that the question of explicit conditions for block positivity remains open.

%%%%%%%%%%%%%%%%%%%%  section 4  %%%%%%%%%%%%%%%%%%%

\section{A three--parameter family of block positive matrices}
\label{secabc}

Fortunately, there exist some particular cases for which positivity conditions \eqref{minorspos} turn out to be useful
 in checking block positivity. Let $a,b,c\in\setC$.
Consider the following family of matrices,
\begin{equation}\label{abcfamilydef}
 F=\matrfour{F}=\left[\begin{array}{cccc}
               	\frac{1}{2}&a&0&0\\
		\bar a&\frac{1}{2}&b&0\\
		0&\bar b&\frac{1}{2}&c\\
		0&0&\bar c&\frac{1}{2}\\
              \end{array}\right],
\end{equation}
which represent operators on $\mathcal{H}_1\otimes\mathcal{H}_2=\setC^2\otimes\setC^2$.

We are going to test condition \eqref{blockposCdef} using the method suggested in the previous section.
 The blocks  (\ref{blocks})
with respect to the subsystem described by ${\cal H}_2$ are
\begin{equation}
 \label{blocksabc}
  F^{\left(2\right)}_u\left(a,b,c\right)=\left[
  \begin{array}{cc}
    \frac{1}{2}\left(\left|u_1\right|^2+\left|u_2\right|^2\right)&a\left|u_1\right|^2+c\left|u_2\right|^2+\bar bu_1\bar u_2\\
    \bar a\left|u_1\right|^2+\bar c\left|u_2\right|^2+b\bar u_1 u_2&\frac{1}{2}\left(\left|u_1\right|^2+\left|u_2\right|^2\right)\\
  \end{array}
 \right].
\end{equation}
Obviously, $F^{\left(2\right)}_u\left(a,b,c\right)$ is semipositive definite for all $u\in\setC^2$ if and 
only if $\det F^{\left(2\right)}_u\left(a,b,c\right)\geqslant 0\forall_{u\in\setC^2}$. That is,
\begin{equation}\label{conditionabc1}
 \left(\frac{1}{2}\left(\left|u_1\right|^2+\left|u_2\right|^2\right)\right)^2-\left|a\left|u_1\right|^2+c\left|u_2\right|^2+
\bar bu_1\bar u_2\right|^2\geqslant 0\haskip\forall_{u_1,u_2\in\setC}.
\end{equation}
Keeping $\left|u_1\right|$ and $\left|u_2\right|$ constant, we can maximize the absolute value of the term 
$a\left|u_1\right|^2+c\left|u_2\right|^2+\bar bu_1\bar u_2$ by choosing the phases of $u_1$ and $u_2$ 
such that the phase of $\bar bu_1\bar u_2$ is the same as the phase of $a\left|u_1\right|^2+c\left|u_2\right|^2$. 
So done, we see that condition \eqref{conditionabc1} is equivalent to
\begin{equation}\label{conditionabc2}
 \left(\frac{1}{2}\left(x^2+y^2\right)\right)^2-\left(\left|ax^2+cy^2\right|+\left| b\right|xy\right)^2\geqslant 0\haskip\forall_{x,y\in\setR^+}.
\end{equation}
In inequality \eqref{conditionabc2}, we substituted $x$ for $\left|u_1\right|$ and $y$ for $\left|u_2\right|$. 
It is now easy to see that \eqref{conditionabc2} is the same as
\begin{equation}\label{conditionabc3}
 \frac{1}{2}\left(x^2+y^2\right)-\left|ax^2+cy^2\right|-\left| b\right|xy\geqslant 0\haskip\forall_{x,y\in\setR}.
\end{equation}
We extended the domain of $x,y$ in \eqref{conditionabc3} to $\setR$, which is permissible
 because $\left| b\right|xy$ does not increase if we change the sign of $x$ or $y$ from plus to minus. 
Substituting $x\rightarrow r\cos\frac{\varphi}{2}$, $y\rightarrow r\sin\frac{\varphi}{2}$ in \eqref{conditionabc3} we obtain 
\begin{equation}
  \label{conditionabc4}
   1-\left|\alpha+\gamma\cos\varphi\right|-\left|b\right|\sin\varphi\geqslant 0\haskip\forall_{\varphi\in\setR},
 \end{equation}
where $\alpha:=a+c$, $\gamma:=a-c$. Condition \eqref{conditionabc4} can be easily solved in the two following situations: 
\begin{itemize}
\item[a)] $\Repart{\alpha\bar\gamma}=0 \ \Longleftrightarrow \  \left|a\right|=\left|c\right|$
\item[b)] $\Repart{\alpha\bar\gamma}=\pm\left|\alpha\right|\left|\gamma\right| \  \Longleftrightarrow \ a=rc,\, r\in\setR$
\end{itemize}
In the case a), condition \eqref{conditionabc4} simplifies to
\begin{equation}
  \label{conditionsita}
 1-\sqrt{\left|\alpha\right|^2+\left|\gamma\right|^2\cos^2\varphi}-\left|b\right|\sin\varphi\geqslant 0\haskip\forall_{\varphi\in\setR}.
\end{equation}
We observe that $\left|\alpha\right|^2+\left|\gamma\right|^2\leqslant 1$ must hold in order that \eqref{conditionsita} be true.
 Keeping this in mind, we can rewrite \eqref{conditionsita} as
\begin{equation}\label{conditionsita2}
 \left|\frac{b}{\gamma}\right|^2\lambda^2-\lambda+\left(1-\left|\frac{b}{\gamma}\right|^2\left(\left|\alpha\right|^2+
\left|\gamma\right|^2\right)\right)\geqslant 0\haskip\forall_{\lambda\in\left[\left|\alpha\right|,\sqrt{\left|\alpha\right|^2
+\left|\gamma\right|^2}\right]}.
\end{equation}
where we substituted $\sqrt{\left|\alpha\right|^2+\left|\gamma\right|^2\cos^2\varphi}\rightarrow\lambda$. 
As a positivity condition for a quadratic function, \eqref{conditionsita2} can be easily solved explicitly.
 Together with the condition on $\left|\alpha\right|^2+\left|\gamma\right|^2$, we obtain
\begin{equation}
\label{conditionsitaexpl}
 \left|\alpha\right|^2+\left|\gamma\right|^2\leqslant 1\land\left|\alpha\right|+
 \left|b\right|^2\leqslant 1\land\left\{2\left|b\right|^2\left|\alpha\right|\leqslant\left|\gamma\right|^2\lor
  2\left|b\right|^2\sqrt{\left|\alpha\right|^2+\left|\gamma\right|^2}\geqslant\left|\gamma\right|^2\right\}
\end{equation}
which is an equivalent form of \eqref{conditionsita2}. 
In the case b), it is even simpler to get the conditions on $\alpha$, $\gamma$ and $b$ equivalent to \eqref{conditionabc4}. We have $\left|\alpha+\gamma\cos\varphi\right|\leqslant\left|\alpha\right|+\left|\gamma\right|\left|\cos\varphi\right|$. Either for $\varphi$ or for $\varphi\rightarrow\pi-\varphi$, we obtain $\left|\alpha+\gamma\cos\varphi\right|=\left|\alpha\right|+\left|\gamma\right|\left|\cos\varphi\right|$ and $\sin\varphi$ is not changed by the substitution $\varphi\rightarrow\pi-\varphi$. Hence we can rewrite \eqref{conditionabc4} as
\begin{equation}\label{casebconditions1}
 1-\left|\alpha\right|-\left|\gamma\right|\left|\cos\varphi\right|-\left|b\right|\sin\varphi\geqslant 0\haskip\forall_{\varphi\in\setR}.
\end{equation}
This is equivalent to $\left(1-\left|\alpha\right|-\left|\gamma\right|\cos\varphi-\left|b\right|\sin\varphi\right)\geqslant 0\forall_{\varphi\in\setR}$, which is easy to solve explicitly in terms of $\alpha$, $\gamma$ and $b$. We get
\begin{equation}\label{casebconditions2}
 1-\left|\alpha\right|-\sqrt{\left|\gamma\right|^2+\left|b\right|^2}\geqslant 0.
\end{equation}
In the case of general $a$, $b$ and $c$, condition \eqref{conditionabc4} is equivalent to
  the following system of four inequalities, 
\begin{eqnarray}
\label{conditionabc6}
{\rm i)} &\Abs{\alpha}+\Abs{\gamma}\  \leqslant  \ 1, \\
\label{conditionabc6b}
{\rm ii)} &\Abs{\gamma}^2-\Abs{b}^2 \ \leqslant  \  \Abs{\Repart{\alpha\bar\gamma}}, \\
\label{conditionabc6c}
{\rm iii)} & \Abs{1-\Abs{\alpha}^2-\Abs{\gamma}^2}\  \geqslant \  2\Abs{\Repart{\alpha\bar\gamma}}, \\
\label{conditionabc5}
{\rm iv)} & \left(\left|\gamma\right|^2+\left|b\right|^2\right)^2\cos^4\psi-4\left(\left|\gamma\right|^2+
\left|b\right|^2\right)\Repart{\alpha\bar\gamma}\cos^3\psi+\\
&+\left(4\Repart{\alpha\bar\gamma}^2-2\left(1-\left|\alpha\right|^2-\left|b\right|^2\right)
\left(\left|\gamma\right|^2+\left|b\right|^2\right)
-4\left|\gamma\right|^2\right)\cos^2\psi+\nonumber\\
&-4\left(3-\left|\alpha\right|^2-\left|b\right|^2\right)\Repart{\alpha\bar\gamma}\cos\psi+\left(1-\left|\alpha\right|^2-
\left|b\right|^2\right)^2-4\left|\alpha\right|^2\  \geqslant  \ 0 \ ,
%\haskip\forall_{\psi\in\setR},
\nonumber
\end{eqnarray}
which have to be satisfied for all real $\psi=2\varphi$. 
The expression on the left hand side of condition iv)
      %\eqref{conditionabc5} 
is a polynomial $P\left(\cos\psi\right)$ of degree four in the variable $\cos\psi$. 
This condition means that $P$ is nonnegative in the interval $\left[-1,1\right]$. 
Given particular values of $a$, $b$ and $c$, 
nonnegativity of $P$ in $\left[-1,1\right]$ can be easily checked 
using the Sturm sequences \cite{Marten}. It is also possible to produce general conditions on $a$, $b$, $c$ in this way, 
but the resulting formulas would be too complicated to reproduce them here and not suitable for further analysis. 

An analogous problem of block positivity over $\setR$
can also be solved for the family of matrices \eqref{abcfamilydef}.
%%%%%%%%
Most of the work has already been done above. We only need to observe that
 the passage from \eqref{conditionabc1} to \eqref{conditionabc2} is possible 
also when $a$, $b$, $c$ and $u_1$, $u_2$ are real numbers. This is true 
because the maximal value of $\Abs{a\left|u_1\right|^2+c\left|u_2\right|^2+bu_1u_2}$ for
fixed $\Abs{u_1}$, $\Abs{u_2}$ is $\Abs{au_1^2+cu_2^2}+\Abs{b}\Abs{u_1}\Abs{u_2}$. Thus the 
condition \eqref{conditionabc2} turns out to be equivalent to block positivity over $\setR$ of 
the matrices of the form \eqref{abcfamilydef} with $a,b,c\in\setR$.
 Later analysis follows as in the case b) discussed above. In this way we arrive at two important conclusions.
 Firstly, symmetric matrices of the form \eqref{abcfamilydef} are block positive over $\setR$
 if and only if they are block positive over $\setC$. 
Secondly, the block positivity condition takes the form \eqref{casebconditions2} with $\alpha=a+c$, $\gamma=a-c$. 
On the other hand, positivity conditions for the family of matrices \eqref{abcfamilydef} are easily obtained,
\begin{equation}
\label{cushionpos}
 \frac{1}{16}-\frac{\left|a\right|^2}{4}-\frac{\left|b\right|^2}{4}-\frac{\left|c\right|^2}{4}+\left|a\right|^2\left|c\right|^2\geqslant 0\land\frac{1}{2}-\left|a\right|^2-\left|b\right|^2-\left|c\right|^2\geqslant 0.
\end{equation}
We can compare them with the block positivity condition \eqref{casebconditions2} in a picture.

%%%%%%%%%%%%%%%   Figure 1 %%%%
%%%%%%%%
\begin{figure}[ph]
  \begin{center}
    \includegraphics[scale=0.55]{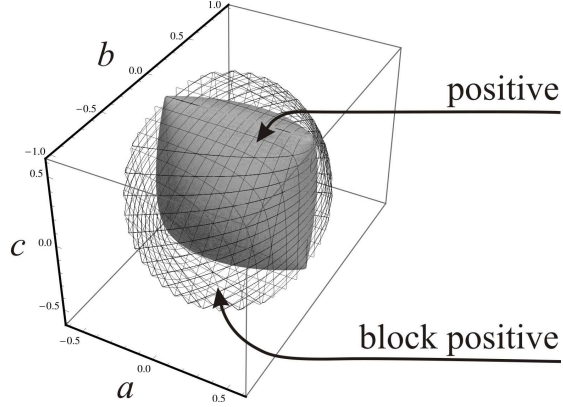}
  \end{center}
  \caption{The grey set of  positive semidefinite matrices
     defined by eq.\,\eqref{cushionpos} is contained inside the set of block positive matrices
    determined by  \eqref{casebconditions2}.
   In this case the block positivity over  $\setC$ is equivalent to the block positivity over $\setR$.
 It is assumed here that $a,b,c\in\setR$, but formulas \eqref{casebconditions1} and 
  \eqref{casebconditions2} apply also for $a,b,c$ complex, provided that $a=r c$ with $r\in\setR$.} 
  \label{fig:author}
\end{figure}
It is clear from Figure \ref{fig:author} that the conditions \eqref{cushionpos} and \eqref{casebconditions2} are not equivalent, 
and  the set of positive matrices of the family \eqref{abcfamilydef}
forms a proper subset of the set of block positive matrices.

A similar investigation can be performed for a related family of matrices,
\begin{equation}\label{abcdfamilydef}
 E\left(s,p,q,r\right)=\left[\begin{array}{cccc}
               	\frac{1}{2}&s&0&r\\
		s&\frac{1}{2}&p&0\\
		0& p&\frac{1}{2}&q\\
		 r&0& q&\frac{1}{2}\\
              \end{array}\right]
\end{equation}
 with real parameters  $s,p,q$ and $r$.
 The block positivity conditions for $E\left(s,p,q,r\right)$ 
 can be obtained using the methods presented in this section. In particular, taking  $E\left(a,\frac{b}{2},c,\frac{b}{2}\right)$ with $a,b,c$ real, 
we get a symmetrization of the family \eqref{abcfamilydef},
\begin{equation}
\label{PTsymmabc}
F'\left(a,b,c\right):=\frac{F\left(a,b,c\right)+F\left(a,b,c\right)^{\tau}}{2}= \left[\begin{array}{cccc}
               	\frac{1}{2}&a&0&\frac{b}{2}\\
		a&\frac{1}{2}&\frac{b}{2}&0\\
		0&\frac{b}{2}&\frac{1}{2}&c\\
		\frac{b}{2}&0&c&\frac{1}{2}\\
              \end{array}\right].
%\frac{F\left(a,b,c\right)+F\left(a,b,c\right)^{\tau}}{2}=
\end{equation}
Deriving conditions for positivity and block positivity of the matrices  
$F'(a,b,c)$, it turns out that in this case both properties do coincide, unlike in the example discussed above.
In the light of Proposition \ref{indecompprop}, this fact can be understood 
as a consequence of the following theorem \cite{Calderon},
\begin{theorem}[Calder\'on]\label{thm.Calderon}Let $m\in\setN$, $x^1,x^2\in\setR$, $\seq{y^j}{j=1}{m},\seq{A_{abcd}}{a,b,c,d}{}\subset\setR$
\begin{equation}\label{formCalderon}
 A_{abcd}x^ay^bx^cy^d\geqslant 0\forall_{x^1,x^2,\seq{y^j}{j=1}{m}}\Longrightarrow A_{abcd}x^ay^bx^cy^d=\sum_i\left(B^i_{ab}x^ay^b\right)^2.
\end{equation}
That is, any positive biquadratic form in $2\times m$ variables is the sum of squares of quadratic forms\qed
\end{theorem}
According to Proposition \ref{indecompprop} and the Calder\'on's result, 
\emph{any operator $A$ on $X\otimes Y\cong\setR^2\otimes\setR^m$ which is symmetric with 
respect to partial transpose and block positive over $\setR$, is decomposable as well}. More than that, 
we know from the discussion preceding Proposition \ref{indecompprop} that $A$ can be written in the 
form $\left(B+B^{\tau}\right)/2$ with $B\geqslant 0$. In the case of $A=F'\left(a,b,c\right)$, $B$ 
must be of the form \eqref{abcdfamilydef} with $s=a$, $q=b$ and $p+r=b$. As can be checked by direct computation, the characteristic polynomials of $E\left(s,p,q,r\right)$ and $E\left(s,r,q,p\right)$ are the same. It follows that $E\geqslant 0\Leftrightarrow E^{\tau}\geqslant 0$, 
which in turn leads us to the conclusion that the matrix $F'=\frac{1}{2}\left(F+F^{\tau}\right)=\frac{1}{2}\left(E+E^{\tau}\right)$ is block positive if 
and only if it is positive.

To explain our observation about $F'$, we could also have used 
the St{\o}rmer-Woronowicz theorem \cite{StormerWor,WorStormer}, 
which implies that \emph{an operator $A$ on $\setR^2\otimes\setR^2$ (or on $\setR^2\otimes\setR^3$)
 is block positive if and only if it is decomposable}. 
This suggests a possible connection between the Calder\'on's and the St\o rmer-Woronowicz theorems.
 On the other hand, the first theorem holds for all $\setR^2\otimes\setR^m$ ($m\in\setN$) whereas the
 latter works for $m\le 3$ only.

The theorem of Calder\'on allows us to find some further implications
for the subject of positive maps.

\begin{proposition}
\label{thm.either}
Let $m\in\setN$.
 Either all block positive operators on $\setC^2\otimes\setC^m$ with \textbf{real matrices} are \textbf{decomposable} or there exists an operator $A$ on $\setC^2\otimes\setC^m$ with real matrix elements $A_{ab,cd}$ such that $A_{ab,cd}x^ay^bx^cy^d$ is the \textbf{sum of squares} of bilinear forms, but $A$ is \textbf{not decomposable}.
\begin{proof}
Let $A$ be an operator on $\setC^2\otimes\setC^m$ with real matrix elements. If $A$ is block positive
 on $\setC^2\otimes\setC^m$, it must be block positive on $\setR^2\otimes\setR^m$. From Calder\' on's theorem 
it follows that $A_{ab,cd}x^ay^bx^cy^d$ is the sum of squares of bilinear forms. If this implies decomposability of $A$,
 any block positive operator on $\setC^2\otimes\setC^m$ with real matrix elements is decomposable. If not, there exists
 an indecomposable operator $A$ such that $A_{ab,cd}x^ay^bx^cy^d$ is SOS.
\end{proof}
\end{proposition}
Both the mutually exclusive possibilities in Proposition \ref{thm.either} are interesting and it 
will be good to know which of them is true for which $m$ (of course, the answer is known for 
$m=1,2,3$ - every positive map is decomposable).
We hope that stronger results of similar kind can also be obtained and they should give
 better insights into the structure of positive and indecomposable maps.

%%%%%%%%%%  section 5  %%%%%%%%%%%

\section{Block positivity of $4\times 4$ matrices over $\setR$}
\label{sec4x4test}
%%%%%%%%%%%%%%%%%%%%%

We want to illustrate the abstract discussion presented in Section \ref{sectestingBP} with a concrete example. 
To that aim, following \cite{pracamag}, 
we derive sufficient and necessary conditions for an arbitrary operator $A$ on $\setR^2\otimes\setR^2$ to
 be block positive 
Let the matrix elements of $A$ be $A_{ab,cd}$ ($a,b,c,d\in\seq{1,2}{}{}$). The blocks with respect to the first subsystem have the matrix elements $\left(A_y^{\left(1\right)}\right)_{ac}=A_{a1,c1}\left(y^1\right)^2+\left(A_{a1,c2}+A_{a2,c1}\right)y^1y^2+A_{a2,c2}\left(y^2\right)^2$. Positivity of $A_y^{\left(1\right)}$ is equivalent to the requirements that $\Tr A_y^{\left(1\right)}\geqslant 0$ and $\det A_y^{\left(1\right)}\geqslant 0$. Nonnegativity of the trace of $A_y^{\left(1\right)}$ for all $y=\left(y^1,y^2\right)\in\setR^2$ means that
\begin{equation}\label{nonnegtr}
\sum_{i=1}^2A_{i1,i1}\left(y^1\right)^2+\sum_{j=1}^2\left(A_{j1,j2}+A_{j2,j1}\right)y^1y^2+\sum_{k=1}^2A_{k2,k2}\left(y^2\right)^2\geqslant 0\haskip\forall_{y^1,y^2\in\setR}.
\end{equation}
Obviously, \eqref{nonnegtr} is a positivity condition for a quadratic form on $\setR^2$ and we can write it explicitly as
\begin{equation}\label{nonnegtrexpl}
 \sum_{i,j=1}^2A_{ij,ij}\geqslant 0\land\sum_{i=1}^2A_{i1,i1}\sum_{k=1}^2A_{k2,k2}-\frac{1}{4}\left(\sum_{j=1}^2\left(A_{j1,j2}+A_{j2,j1}\right)\right)^2\geqslant 0.
\end{equation}
The expression for the determinant of $A_y^{\left(1\right)}$ reads
\begin{equation}\label{polord4blockposRhomogeno}
 \det A_y^{\left(1\right)}=c_4 x^4+c_3 x^3z+c_2 x^2z^2+c_1 xz^3+c_0z^4,
\end{equation}
where we substituted $x$ for $y^1$, $z$ for $y^2$ and we introduced 
\begin{eqnarray}\label{cis}
c_0=&A_{12,12}A_{21,21}-A_{11,21}A_{21,11},\label{cis0}\\
c_1=&A_{22,22}\left(A_{12,11}+A_{11,12}\right)+A_{12,12}\left(A_{22,21}+A_{21,22}\right)+\label{cis1}\\&-A_{22,12}\left(A_{12,21}+A_{11,22}\right)-A_{11,21}\left(A_{22,11}+A_{21,12}\right),\nonumber\\
c_2=&A_{11,11}A_{22,22}+A_{21,21}A_{12,12}+\left(A_{11,12}+A_{12,11}\right)\left(A_{21,22}+A_{22,21}\right)+\label{cis2}\\
&-A_{11,21}A_{21,11}-A_{12,22}A_{22,12}-\left(A_{11,22}+A_{12,21}\right)\left(A_{21,12}+A_{22,11}\right),\nonumber\\
c_3=&A_{11,11}\left(A_{21,22}+A_{22,21}\right)+\left(A_{11,12}+A_{12,11}\right)+\label{cis3}\\
&-A_{11,21}\left(A_{21,12}+A_{22,11}\right)-A_{22,12}\left(A_{11,22}+A_{12,21}\right)\nonumber,\\
c_4=&A_{11,11}A_{21,21}-A_{11,21}A_{21,11}.\label{cis4}
\end{eqnarray}
The $c_i$'s are homogeneous polynomials in the matrix elements $A_{ab,cd}$. 
It is easy to see that non-negativity of \eqref{polord4blockposRhomogeno} for all $x,z\in\setR$ is equivalent to
\begin{equation}\label{polord4blockposR}
 c_4 x^4+c_3 x^3+c_2 x^2+c_1 x+c_0\geqslant 0\haskip\forall_{x\in\setR}.
\end{equation}

Thus we showed that in the case of a symmetric matrix $A$ of order $4$
condition \eqref{blockposdefRind} is equivalent to \eqref{nonnegtrexpl} plus \eqref{polord4blockposR}. 
The inequalities \eqref{nonnegtrexpl} are explicit conditions on the matrix elements $A_{ab,cd}$, 
but in \eqref{polord4blockposR} we need some additional work to dispose of the quantifier $\forall_{x\in\setR}$. 
There is no single method of doing it, but the one which seems most economical to us is by using the following 
theorem \cite{Turowicz},
\begin{theorem}[Sturm]\label{thm.Sturm}Let $f=f_0$ be a real univariate polynomial with no multiple roots in $\setR$. Let $f_1$ be the first derivative of $f$. Define
\begin{equation}\label{fml.defoffn.1}
f_{n+1}:=\fnc{\textnormal{rem}}{f_{n-1},f_n},
\end{equation}
where $\fnc{\textnormal{rem}}{h,g}$ is the remainder obtained when dividing $h$ by $g$. Define $\fnc{N}{r}$ as the number of sign changes in the sequence
\begin{equation}\label{fml.fnseq.1}
 \fnc{f_0}{r},\fnc{f_1}{r},-\fnc{f_2}{r},-\fnc{f_3}{r},\fnc{f_4}{r},\fnc{f_5}{r},-\fnc{f_6}{r},\ldots
\end{equation}
with zeros skipped. Assume $\alpha,\beta\in\setR$, $\alpha<\beta$, $\fnc{f_0}{\alpha}\neq 0$ and $\fnc{f_0}{\beta}$. The number of zeros of $f_0$ in the interval $\left(\alpha,\beta\right)$ equals $\fnc{N}{\alpha}-\fnc{N}{\beta}$,
\begin{equation}\label{fnc.Sturmthm.1}
\fnc{N}{\alpha}-\fnc{N}{\beta}=\#\left\{r\in\left(\alpha;\beta\right)|\fnc{f}{r}=0\right\}.
\end{equation}\qed
\end{theorem}
The sequence of functions \eqref{fml.defoffn.1} is the same as in the Euclid's algorithm applied to $f$ and $f'$. When the signs are changed as in \eqref{fml.fnseq.1}, the sequence is called the \textit{Sturm sequence} of $f$. We know that $\left\{f_n\right\}_{n=0,1\ldots}$ must terminate at some $f_m\in\setR\setminus\seq{0}{}{}$, which is the greatest common divisor of $f$ and $f'$. If we go to the limits $\alpha=-\infty$, $\beta=+\infty$ in Theorem \ref{thm.Sturm}, we easily obtain the number of real roots of $f$,
\begin{corollary}
\label{cor.Sturminfty}
 Let $f=f_0$ be a real univariate polynomial with no multiple roots in $\setR$ and $f_1$ - its first derivative. Let $f_n$ \textnormal{($n=2,3\ldots$)} be defined like in \fml{defoffn.1} and assume
\begin{equation}\label{fml.coeffsoffn.1}
 \fnc{f_n}{r}=a_{n,k_n}r^{k_n}+a_{n,k_n-1}r^{k_n-1}+\ldots+a_{0,n},
\end{equation}
where $k_n\geqslant 0$, $a_{n,k_n}\neq 0$ $\forall_n$. Denote with $\fnc{N}{+\infty}$ the number of sign changes in the sequence
\begin{equation}\label{fml.fnseq.2}
a_{0,k_0},a_{1,k_1},-a_{2,k_2},-a_{3,k_3},a_{4,k_4},\ldots,\pm a_{m,k_m}
\end{equation}
and with $\fnc{N}{-\infty}$ the number of sign changes in
\begin{equation}\label{fml.fnseq.3}
\left(-\right)^{k_0}a_{0,k_0},\left(-\right)^{k_1}a_{1,k_1},\left(-\right)^{k_2+1}a_{2,k_2},\left(-\right)^{k_3+1}a_{3,k_3},\left(-\right)^{k_4}a_{4,k_4},\ldots,\pm a_{m,k_m}.
\end{equation}
The number of real zeros of $f$ equals $\fnc{N}{-\infty}-\fnc{N}{+\infty}$,
\begin{equation}\label{fnc.Sturmthm.2}
 \fnc{N}{-\infty}-\fnc{N}{+\infty}=\#\left\{x\in\setR|\fnc{f}{x}=0\right\}.
\end{equation}\qed
\end{corollary}
Let us take for $f$ the polynomial $c_4 x^4+c_3 x^3+c_2 x^2+c_1 x+c_0$ which appears in \eqref{polord4blockposR}. We shall now assume that it has no multiple roots in $\setR$. Then we can use Corollary \ref{cor.Sturminfty} to check positivity of $f$. 

The sequence \SEQ{f_n}{n=0,1,\ldots}{} consists of at most five polynomials,
\begin{eqnarray}\label{fml.fnseqforord4.0}
 f = f_0 =& c_4 x^4 +c_3 x^3 +c_2x^2 + c_1x + c_0,\\
 f_1 =& 4c_4x^3+3c_3x^2+2c_2x+c_1,\nonumber\\
 f_2 =& a_{2,2} x^2+a_{2,1} x+a_{2,0},\nonumber\\
 f_3 =& a_{3,1} x +a_{3,0},\nonumber\\
 f_4 =& a_{4,0}.\nonumber
\end{eqnarray}
If we make an additional \textit{normality} assumption, which says that the degrees of $f_0,\ldots,f_4$ drop one by one in the successive lines of \eqref{fml.fnseqforord4.0}, it is easy to write down positivity conditions for $f$,
\begin{equation}\label{poscond4abcd}
 c_4>0\land\left(a_{2,2}>0\lor a_{3,1}>0\right)\land a_{4,0}>0.
\end{equation}
The expressions for $a_{2,2}$, $a_{3,1}$ and $a_{4,0}$ can also be easily obtained in the present situation. We get
\begin{equation}\label{expressionscoeffs}
 a_{2,2}=\frac{1}{16c_4}\sigma_1,\quad a_{3,1}=\frac{32c_4}{\sigma_1^2}\sigma_2,\quad a_{4,0}=-\frac{\sigma_1^2}{64c_4\sigma_2^2}\sigma_3,
\end{equation}
where
\begin{eqnarray}\label{sigmasi}
\sigma_1:=&\hskip -14 pt 8 c_2 c_4-3 c_3^2,\label{sigmasi1}\\
\sigma_2:=&\hskip -14 pt 3 c_1 c_3^3-14 c_1 c_2 c_3 c_4-c_3^2 \left(c_2^2-6 c_0 c_4\right)+2 c_4
\left(2 c_2^3+9 c_1^2 c_4-8 c_0 c_2 c_4\right)\label{sigmasi2},\\
\sigma_3:=&\hskip -14 pt 2 {c_1} \left(2 {c_1}^2-9 {c_0} {c_2}\right) {c_3}^3+2 {c_1} {c_3} {c_4} \left(-9
{c_1}^2 {c_2}+40 {c_0} {c_2}^2+96 {c_0}^2 {c_4}\right)+\label{sigmasi3}\\
&\hskip -14 pt 27 {c_0}^2 {c_3}^4+{c_3}^2 \left(-{c_1}^2 {c_2}^2+4 {c_0} {c_2}^3+6
{c_0} {c_1}^2 {c_4}-144 {c_0}^2 {c_2} {c_4}\right)+\nonumber\\
&\hskip -14 pt {c_4} \left(4 {c_1}^2 {c_2}^3+27 {c_1}^4 {c_4}+128
{c_0}^2 {c_2}^2 {c_4}-256 {c_0}^3 {c_4}^2-16 {c_0} {c_2} \left({c_2}^3+9 {c_1}^2 {c_4}\right)\right).\nonumber
\end{eqnarray}
According to \eqref{expressionscoeffs}, the normality assumption is equivalent to $c_4\neq 0\land\sigma_1\neq 0\land\sigma_2\neq 0\land\sigma_3\neq 0$. If these conditions hold, we can rewrite \eqref{poscond4abcd} as
\begin{equation}
\label{poscond4sigmas}
 c_4>0 \ \land \ \left(\sigma_1>0\lor\sigma_2>0\right)\ \land\ \sigma_3<0.
\end{equation}
This is an explicit condition for $f$ to be positive. Of course, it was obtained under the assumption 
that \SEQ{f_n}{n=0}{4} are normal. Nevertheless, we can use \eqref{poscond4sigmas} as a starting 
point for an all-purpose nonnegativity test for polynomials of degree less or equal four. Indeed, suppose that the sequence \SEQ{f_n}{n=0}{4} is not normal. That is, at least one of the numbers $c_4,\sigma_1,\sigma_2,\sigma_3$ happens to be zero. If $c_4=0$, the nonnegativity question becomes trivial. We get that $c_3 x^3+c_2 x^2+c_1 x+c_0$ is nonnegative if and only if
\begin{equation}
\label{poscondquadfnc}
 c_3=0 \ \land \ c_2\geqslant 0\  \land \  c_1^2-4c_2c_0\leqslant 0.
\end{equation}
The case in which  $c_4\neq 0$ but $\sigma_1\sigma_2\sigma_3=0$ can be analyzed using a little more sophisticated techniques (see Appendix \ref{apppol4B}). All in all, we arrive at the following nonnegativity conditions for $f$,
\begin{multline}
\label{nonnegdeg4}
 \left\{c_4>0\ \land\ \left(\sigma_1\geqslant 0\lor\sigma_2\geqslant 0\right)\ \land\ \sigma_3\dot<0\right\}\lor\\
\lor\left\{c_4=0\land c_3=0\land c_2\geqslant 0\land c_1^2-4c_2c_0\leqslant 0\right\},
\end{multline}
where $\sigma_3\dot<0$ means 
$\exists_{\xi>0}\forall_{\xi'<\xi}\left(\sigma_3\left(c_4,c_3,c_2,c_1,c_0+\xi'\right)<0\right)$. We can write $\sigma_3\dot<0$ explicitly as
\begin{equation}
\label{sigmadotlt0}
 \sigma_3<0\ \lor\ \Bigl(\sigma_3=0\ \land\ \bigl(\kappa_1<0\ \lor\ 
             (\kappa_1=0\ \land\ \kappa_2\leqslant0 ) \bigr) \Bigr) \ ,
\end{equation}
where
\begin{eqnarray}\label{kappy}
 \kappa_1=4 c_3^2 c^3-18 c_3^3 c_2 c_1+80 c_4 c_3 c_2^2 c_1+6 c_4 c_3^2 c_1^2-16 c_4 c_2 \left(c_2^3+9 c_4 c_1^2\right)\label{kappa1},\\
\kappa_2=27 c_3^4-144 c_4 c_3^2 c_2+128 c_4^2 c_2^2+192 c_4^2 c_3 c_1.\label{kappa2}
%\kappa_3=-256 c_4^3\label{kappa3}
\end{eqnarray}
Obviously,  conditions \eqref{nonnegdeg4} and \eqref{nonnegtrexpl} together with the definitions 
\eqref{sigmadotlt0}, \eqref{kappa1}, \eqref{kappa2}, \eqref{sigmasi1}, \eqref{sigmasi2}, \eqref{sigmasi3},
  \eqref{cis0}, \eqref{cis1}, \eqref{cis2}, \eqref{cis3} and \eqref{cis4} provide us with a method to test 
block positivity over $\setR$ of $4\times 4$ matrices. We see that lengthy
 calculations are involved, even though the studied example is the simplest possible one. 
It is also clear that the iterative procedure proposed in \cite{Jamiolkowski1} could not work with conditions
 like \eqref{poscond4sigmas}, let alone \eqref{nonnegdeg4}.

\section{Conclusions}
We have re-examined the method \cite{Jamiolkowski1} of establishing positivity of a map 
with help of multivariate polynomials
and we conclude that in the general case this problem  remains open.
The same can be said about the equivalent problem of checking whether a given operator acting on a composite Hilbert space is block positive.
%It is worth to stress that this problem is equivalent to
%checking the block positivity of the corresponding operator acting on a composite %Hilbert space.
%
% The method for solving this problem suggested in \cite{Jamiolkowski1} is not conclusive, 
% as we explained in Section \ref{sectestingBP}. 
% \footnote{that was also one of the main reasons for writing Propositions 
% \ref{indecompprop} and \ref{thm.either}} between block positivity over $\setC$ and over $\setR$. 
% This subject needs further studies.%\textit{+the need to explore the relation between block positivity 
% over $\setC$ and over $\setR$}
%
%On the other hand,
Nevertheless, for certain family of operators checking the positivity of the 
associated polynomials allowed us to find concrete criterion for block positivity.
Such concrete examples are provided in Sections \ref{secabc} and \ref{sec4x4test}.
By giving the example \eqref{exmplbpCnebpR}, we touched upon the relation between the block positivity conditions 
over $\setC$ and over $\setR$.

We also outlined connections between block positivity, indecomposability 
and the sums of squares (Propositions \ref{indecompprop} and \ref{thm.either}, 
Theorem \ref{thm.Calderon}). Proposition \ref{thm.either} opens a discussion about the 
two mutually exclusive possibilities concerning indecomposable maps 
on $\setC^2\otimes\setC^m$ (cf. Section \ref{secabc}).

Finally, we tried to show that polynomials, which have been thoroughly studied by mathematicians and engineers,
 may deserve more respect of physicists working on quantum information or on open quantum systems. 
In particular, the separability problem itself can be formulated as a set of polynomial equalities \cite{Korbicz}.
 Techniques like the calculation of a Gr\"obner basis of an ideal are widely used to solve polynomial equations 
and they could be of importance in physical problems like the separability problem.

\section{Acknowledgements}
The authors wish to thank Andrzej Jamio{\l}kowski for very useful comments 
and for providing them with copies of his early papers \cite{Jamiolkowski1,Jamiolkowski2}.
 We also acknowledge helpful remarks by Dariusz Chru{\'s}ci{\'n}ski and Pawe{\l} Horodecki.
 The work was financially supported by the special grant number DFG-SFB/38/2007 of
Polish Ministry of Science and Higher Education and the European Research Project  COCOS.
%%
%%%%%%%%%%  APPENDIX         %%%%%%%%%%%%%%%%
%
\appendix
%THIS SECTION WAS WRONG
\section{Equality between operators   $\tilde A$ and $A$}
%%%%%%%%%%%%%%         appendix A        %%%%%%%%%%
\label{apphiA}
Consider the symmetric bilinear forms $\Phi:\left(X\otimes Y\right)^2\ni\left(w_1,w_2\right)\mapsto w_1\cdot A\left(w_2\right)\in\setR$, $\tilde\Phi:\left(X\otimes Y\right)^2\ni\left(w_1,w_2\right)\mapsto w_1\cdot \tilde A\left(w_2\right)\in\setR$. We know that $\Phi\left(x\otimes y\right)=\tilde\Phi\left(x\otimes y\right)$ for arbitrary $x\in X$, $y\in Y$. From \eqref{partTA} and \eqref{Atildematr} it follows that $\Phi$, $\tilde\Phi$ are symmetric with respect to partial transposition,
\begin{eqnarray}
 \Phi\left(x_1\otimes y_1,x_2\otimes y_2\right)=&\Phi\left(x_1\otimes y_2,x_2\otimes y_1\right)\label{PhiPT},\\
\tilde\Phi\left(x_1\otimes y_1,x_2\otimes y_2\right)=&\tilde\Phi\left(x_1\otimes y_2,x_2\otimes y_1\right)\label{tildePhiPT}.
\end{eqnarray}
Let us choose $x\in X$ and consider the following maps, $\Phi_x:Y^2\ni y\mapsto\Phi\left(x\otimes y_1,x\otimes y_2\right)\in\setR$, $\tilde\Phi_x:Y^2\ni y\mapsto\Phi\left(x\otimes y_1,x\otimes y_2\right)\in\setR$. From \eqref{PhiPT} and \eqref{tildePhiPT} we know that $\Phi_x$, $\tilde\Phi_x$ are symmetric bilinear forms on $Y$. As a consequence of $\Phi\left(x\otimes y\right)=\tilde\Phi\left(x\otimes y\right)$, $\Phi_x\left(y,y\right)=\tilde\Phi_x\left(y,y\right)$ for arbitrary $y\in Y$. Hence the quadratic forms corresponding to $\Phi$ and $\tilde\Phi$ are equal. This implies $\Phi_x=\tilde\Phi_x$, so we get
\begin{equation}
\label{PhixtildePhix}
 \Phi\left(x\otimes y_1,x\otimes y_2\right)=\tilde\Phi\left(x\otimes y_1,x\otimes y_2\right)\haskip\forall_{x\in X,y_1,y_2\in Y}.
\end{equation}
Now we consider the maps $\Phi_{y_1,y_2}:X^2\ni\left(x_1,x_2\right)\mapsto\Phi\left(x_1\otimes y_1,x_2\otimes y_2\right)$, $\tilde\Phi_{y_1,y_2}:X^2\ni\left(x_1,x_2\right)\mapsto\tilde\Phi\left(x_1\otimes y_1,x_2\otimes y_2\right)$. From the symmetry of $\Phi$, $\tilde\Phi$ and the properties \eqref{PhiPT}, \eqref{tildePhiPT}, we see that $\Phi_{y_1,y_2}$, $\tilde\Phi_{y_1,y_2}$ are symmetric bilinear forms on $X$. As a consequence of \eqref{PhixtildePhix}, $\Phi_{y_1,y_2}\left(x,x\right)=\tilde\Phi_{y_1,y_2}\left(x,x\right)$ for all $x\in X$. This implies $\Phi_{y_1,y_2}\left(x_1,x_2\right)=\tilde\Phi_{y_1,y_2}\left(x_1,x_2\right)$ for arbitrary $x_1,x_2\in X$. In this way we get $\Phi\left(x_1\otimes y_1,x_2\otimes y_2\right)=\tilde\Phi\left(x_1\otimes y_1,x_2\otimes y_2\right)\forall_{x_1,x_2\in X,y_1,y_2\in Y}$, which is the same as
\begin{equation}\label{equAtildeA}
 \left(x_1\otimes y_1\right)\cdot A\left(x_2\otimes y_2\right)=\left(x_1\otimes y_1\right)\cdot\tilde A\left(x_2\otimes y_2\right)\haskip\forall_{x_1,x_2\in X}\forall_{y_1,y_2\in Y}.
\end{equation}
Of course, \eqref{equAtildeA} implies $A=\tilde A$.
%%
%%%%%%%%%%%%%%         appendix B        %%%%%%%%%%
%%
\section{Nonnegative polynomials with $\sigma_1\sigma_2\sigma_3=0$ and $c_4\neq 0$}
\label{apppol4B}
Our aim is to figure out all the sign configurations of $c_4,\sigma_1,\sigma_2,\sigma_3$ such that they meet 
the constraints $c_4\neq 0\land \sigma_1\sigma_2\sigma_3=0$ and they correspond to nonnegative polynomials
 $f=c_4x^4+c_3x^3+c_2x^2+c_1x^1+c_0$. We also have to check that the remaining sign configurations can never
 give a nonnegative $f$. Of course, $c_4<0$ implies that $f\left(x\right)$ be negative for some $x$, so we only need to 
consider $c_4$ positive. First we show that $\sigma_3>0$ cannot happen for a nonnegative $f$. Suppose $\sigma_3>0$. 
We know that $\sigma_1=0$ or $\sigma_2=0$. Let us first consider $\sigma_1=0$. Because $\sigma_1=8c_2c_4-3c_3^2$ 
and $c_4>0$, we can increase $c_2$ by $\varepsilon>0$ and get $\sigma_1>0$ for sure. If $\sigma_2$ turns out to be 
zero after this operation, we additionally increase $c_0$ by $\xi>0$, which must give us $\sigma_2\neq0$ because 
$\sigma_2=\tilde\sigma_2-2c_0c_4\sigma_1$ where $\tilde\sigma_2$ does not depend on $c_0$. The numbers 
$\varepsilon$, $\xi$ can be made arbitrarily small, so as not to influence the sign of $\sigma_3$. Hence we see that 
$f+\varepsilon x^2+\xi$ has a normal Sturm sequence and it does not satisfy \eqref{poscond4sigmas} because the 
$\sigma_3$ corresponding to $f+\varepsilon x^2+\xi$ is positive. But $f+\varepsilon x^2+\xi\not>0$ implies 
$f\not\geqslant 0$, so $f$ cannot be nonnegative. We conclude that $f\geqslant 0$ is impossible for $c_4>0$, 
$\sigma_3>0$ and $\sigma_1=0$. For $c_4>0$, $\sigma_3>0$ and $\sigma_2=0$, we only need to increase 
$f$ by a sufficiently small $\xi$ to get to the conclusion $f\not\geqslant 0$. Our observations mean 
that $\sigma_3>0$ always implies $f\not\geqslant 0$. 
Let us now consider the polynomials $f$ for which conditions
  \begin{equation}
    \label{condsigm10sigm20}
 c_4>0\ \land\ \left(\sigma_1=0\lor\sigma_2=0\right)\ \land\ \sigma_3<0.
\end{equation}
are satisfied. If $\sigma_1$ vanishes, we can get $\sigma_1>0$ by increasing
 $c_2$ ($c_2\rightarrow c_2+\varepsilon$). If $\sigma_2$ turns out to be $0$ afterwards, 
any change of $c_0$ ($c_0\rightarrow c_0+\xi$) will give us $\sigma_2\neq 0$ (cf. the discussion above). 
We can take $\varepsilon$ and $\xi$ arbitrarily small, which allows us to avoid changing the sign of $\sigma_3$. 
After all, we get a polynomial $f+\varepsilon x^2+\xi$ which has a normal Sturm sequence and it is 
positive since $c_4>0\land\sigma_1>0\land\sigma_3>0$ for the corresponding $c_4$, $\sigma_1$ and $\sigma_3$.
 Because $\varepsilon$ and $\xi$ can be arbitrarily small, we see that $f$ is a pointwise limit of a sequence 
of positive polynomials. Hence $f$ is nonnegative. The same conclusion can be drawn for $c_4>0$, 
$\sigma_1\neq 0$, $\sigma_2=0$ and $\sigma_3<0$, so we should add \eqref{condsigm10sigm20} to our 
set of non-negativity conditions. We can write \eqref{condsigm10sigm20} and \eqref{poscond4sigmas} as a single condition,
\begin{equation}
\label{condg00g}
 c_4>0\  \land\  \left(\sigma_1\geqslant0\lor\sigma_2\geqslant0\right) \ \land\  \sigma_3<0 \  .
\end{equation}
The only situation which is left to analyze is that of $\sigma_3=0$. To that end, let us write $\sigma_3$ as a polynomial in $c_0$,
\begin{equation}\label{sigma3polc0}
 \sigma_3=\kappa_3c_0^3+\kappa_2 c_0^2+\kappa_1c_0+\kappa_0,
\end{equation}
where
\begin{eqnarray}\label{kappy2}
\kappa_0=-c_3^2 c_2^2 c_1^2+4 c_4 c_2^3 c_1^2+4 c_3^3 c_1^3-18 c_4 c_3 c_2 c_1^3+27 c_4^2 c_1^4\label{kappa00},\\
 \kappa_1=4 c_3^2 c^3-18 c_3^3 c_2 c_1+80 c_4 c_3 c_2^2 c_1+6 c_4 c_3^2 c_1^2-16 c_4 c_2 \left(c_2^3+9 c_4 c_1^2\right)\label{kappa11},\\
\kappa_2=27 c_3^4-144 c_4 c_3^2 c_2+128 c_4^2 c_2^2+192 c_4^2 c_3 c_1\label{kappa22},\\
\kappa_3=-256 c_4^3.\label{kappa33}
\end{eqnarray}
Because of the assumption $c_4>0$, we know that $\sigma_3$ is not constant with respect to $c_0$. 
The idea now is to infinitesimally increase $c_0$ and see what the outcome is. If $\sigma_3$ becomes positive, we conclude that $f\not\geqslant 0$. If it turns out to be negative (we denote this with $f\dot<0$), we go back to the initial values of $c_i$ and ask about the signs of $\sigma_1$, $\sigma_2$. If $\sigma_1>0$, we choose $\xi>0$ so small that $\sigma_1>0$ holds when we increase $c_0$ by $\xi$. Then \eqref{condg00g} is true for $f+\xi$, so $f+\xi$ is nonnegative. Since the $\xi$ in $f+\xi$ can be made arbitrarily small, we get $f\geqslant 0$. If $\sigma_1=0$, we increase $c_0$ by $\xi$ to get $\sigma_3<0$ and then we increase $c_2$ by $\varepsilon$ so as to get $\sigma_1>0$ and not to violate $\sigma_3>0$. After that \eqref{condg00g} holds for $f+\varepsilon x^2+\xi$ and again we get to the conclusion that $f\geqslant 0$. Therefore we can add
\begin{equation}
\label{condgg0g}
 c_4>0\ \land\ \sigma_1\geqslant 0\ \land\ \sigma_3\dot<0.
\end{equation}
to our list of non-negativity conditions for $f$. Now we only need to analyze the case $c_4>0\land\sigma_1<0\land\sigma_2\geqslant\land\sigma_3\dot <0$ to finish our work. If $\sigma_2>0$, we choose $\xi>0$ so small that the $\sigma_2$ corresponding to $f+\xi$ is also positive. Then $f+\xi>0$ and we get $f\geqslant 0$. The case $\sigma_2=0$ is also simple to analyze. Because $\sigma_1<0$, increasing $c_0$ causes $\sigma_2=\tilde\sigma_2-c_4c_0\sigma_1$ to become positive, so we get $\sigma_2>0\land\sigma_3<0$ for $f+\xi$ and again this leads us to $f\geqslant 0$. Thus we can add
\begin{equation}\label{condgmg0g}
 c_4>0\ \land\ \sigma_1<0\ \land\ \sigma_2\geqslant 0\ \land\ \sigma_3\dot <0.
\end{equation}
to our nonnegativity conditions for $f$. It is convenient to write \eqref{condg00g} and \eqref{condgg0g} in a single formula,
\begin{equation}\label{condgg0g0g}
 c_4>0\ \land\ \left(\sigma_1\geqslant 0\lor\sigma_2\geqslant 0\right)\ \land\ \sigma_3\dot <0.
\end{equation}
Because of the particular form \eqref{sigma3polc0} of $\sigma_3$, we explicitly write the condition $\sigma_3\dot <0$ as
\begin{equation}
\label{condlt0lt0lt0}
 \sigma_3<0\ \lor\ \Bigl( \sigma_3=0\ \land\ \bigl( \kappa_1<0\ \lor\ ( \kappa_1=0\ \land\ \kappa_2\leqslant0) \bigr) \Bigr)\ ,
\end{equation}
since  condition $c_4>0$ implies that $\kappa_3<0$.


\begin{thebibliography}{4}

\bibitem{StormerWor}E. St\o rmer, \textit{Positive linear maps of operator algebras}, Acta. Math. 110 (1963), 233-278

\bibitem{Jamiolkowski}A. Jamio{\l}kowski, 
\textit{Linear transformations which preserve trace and positive semidefiniteness of operators},
 Rep. Math. Phys. 3 (1972), 275-278

\bibitem{WorStormer}S. L. Woronowicz, \textit{Positive maps of low-dimensional matrix algebra}, 
                               Rep. Math. Phys. 10 (1976), 165-183

\bibitem{MM01} W. A. Majewski and M. Marciniak, 
   \textit{On a characterization of positive maps}, 
     J. Phys. A34  (2001), 5836

\bibitem{SSZ09}  {\L}. Skowronek,  E. St\o rmer, and K. {\.Z}yczkowski,    
\textit{Cones of positive maps and their duality relations},
preprint arXiv:0902.4877

\bibitem{Choi75}M. D. Choi, \textit{Positive Semidefinite Biquadratic Forms}, Linear Alg. Appl. 12 (1975), 95-100

\bibitem{Tomiyama}K. Tanahashi, J. Tomiyama, \textit{Indecomposable positive maps in matrix algebras}, Canad. Math. Bull. 31 (1988), 308-317

\bibitem{HaAtomic}K. C. Ha, \textit{Atomic positive linear maps in matrix algebras}, 
Publ. RIMS, Kyoto Univ. 34 (1998), 591-599

\bibitem{Sudarshan} A. Shaji, E. C. G. Sudarshan, \textit{Who's afraid of not completely positive maps?}, Phys. Lett. A 341 (2005), 48-54

\bibitem{Pachukas}P. Pechukas, \textit{Reduced Dynamics Need Not Be Completely Positive}, Phys. Rev. Lett. 73 (1994), 1060-1062

\bibitem{Buzek}P. \v Stelmanovi\v c, V. Bu\v zek, \textit{Dynamics of open systems initially entangled with environment: Beyond the Kraus representation}, Phys. Rev. A 64 (2001), 062106

\bibitem{Horodeccy}M. Horodecki, P. Horodecki, R. Horodecki, 
\textit{Separability of Mixed States: Necessary and Sufficient Conditions},
 Phys. Lett.  A 223 (1996), 1

\bibitem{TerhalIndec} B. M. Terhal, 
\textit{A Family of Indecomposable Positive Linear Maps based on Entangled Quantum States}, 
Lin. Alg. Appl. 323 (2000), 61-73

\bibitem{LewensteinKraus}M. Lewenstein,  B. Kraus,  P. Horodecki, J. I. Cirac, 
 \textit{Characterization of separable states and entanglement witnesses}, Phys. Rev. A 63 (2001), 044304

\bibitem{ChruscinskiKossakowski1} D. Chru{\'s}ci{\'n}ski, A. Kossakowski, 
\textit{On the structure of entanglement witnesses and new class of positive indecomposable maps}, 
Open Sys. Information Dyn. 14 (2007), 275-294

\bibitem{ChruscinskiKossakowski2} D. Chru{\'s}ci{\'n}ski, A. Kossakowski, 
\textit{Spectral conditions for positive maps}, preprint arXiv:0809.4909

\bibitem{GHB+03} O. G{\"u}hne et al.,
%O. Guehne, P. Hyllus, D. Bru{\ss}, A. Ekert, M. Lewenstein, C. Macchiavello, A. Sanpera,
 \textit{Experimental detection of entanglement via witness operators and local measurements},
J. Mod. Opt. 50, 1079 (2003)

\bibitem{Jamiolkowski1}A. Jamio{\l}kowski, 
\textit{An effective method of investigation of positive maps on the set of positive definite operators}, 
Rep. Math. Phys. 5 (1974), 415-424

\bibitem{Jamiolkowski2}A. Jamio{\l}kowski, \textit{On semipositive definiteness of $2n$-degree forms},
 Rep. Math. Phys. 10 (1976), 259-266


\bibitem{Stormer}E. St\o rmer, \textit{Decomposable Positive Maps on $C^{\ast}$-Algebras}, Proc. Amer. Math. Soc. 86 (1982), 402-404

\bibitem{MichauxOzturk}C. Michaux, A. Ozturk, \textit{Quantifier elimination following Muchnik}, Universit\'e de Mons-Hainaut, Institute de Math\'ematique, Preprint 10, April 11, 2002, {\footnotesize\ttfamily http://math.umh.ac.be/preprints/src/Ozturk020411.pdf}

\bibitem{Marten}M. Marten, \textit{Geometry of polynomials}, AMS Surveys 3, Third edition, Providence, RI (1985)

\bibitem{Calderon}A. P. Calder\'on, \textit{A note on biquadratic forms}, Linear Alg. Appl. 7 (1973), 175-177


\bibitem{Turowicz}A. Turowicz, \textit{Geometria zer wielomian{\'o}w}, PWN, Warszawa, 1967 (in Polish)

\bibitem{pracamag} {\L}. Skowronek, 
\textit{Quantum Entanglement and certain problems in mathematics}, 
Master's Thesis, Krakow, June 2008,
 {\footnotesize\ttfamily http://chaos.if.uj.edu.pl/$\tilde\,$karol/prace/skowronek08.pdf}

\bibitem{Korbicz}J. Korbicz, F. Hulpke, A. Osterloh, M. Lewenstein,
 \textit{A statistical-mechanical description of quantum entanglement}, 
J. Phys. A 41 (2008) 375301

\end{thebibliography}
\end{document}